\documentclass[12pt]{article}

\usepackage{amssymb}
\usepackage{amsmath}

\usepackage{epsfig}
\begin{document}%
%%%%%%%%%%%%%%%%%

%%%%%%%%%%%%%%%%%%%%%%%%%%%%%%%%%%%%%%%%%%%%%%%%%%%%%%%%%%%%%%%%%%%%%

\title{\bf Gravity effects in inclined air showers induced by cosmic neutrinos}

\author{A.V. Kisselev\thanks{Electronic address: alexandre.kisselev@ihep.ru} \\
\small Institute for High Energy Physics, 142281 Protvino, Russia}

\date{}

\maketitle

\thispagestyle{empty}

\bigskip

\begin{abstract}
The Randall-Sundrum  model with a small curvature is considered in
which five-dimensional Planck scale lies in the TeV region, and a
spectrum of Kaluza-Klein gravitons reminds that in one flat extra
dimension. The cross sections for interactions of ultra-high
energy cosmic neutrinos with nucleons are calculated. It is shown
that effects related with massive graviton excitations can be
detected in deeply penetrating inclined air showers induced by
these neutrinos. The expected number of air showers at the Auger
Observatory is estimated as a function of two parameters of the
model.
\end{abstract}

%%%%%%%%%%%%%%%%%%%%%%%%%%%%%%%%%%%%%%%%%%%%%%%%%%%%%%%%%%%%%%%%%%%%%

%%%%%%%%%%%%%
% Main text %
%%%%%%%%%%%%%

\section{Warped extra dimension with the small curvature}

One of the most important problem of the modern particle physics
is a hierarchy problem, i.e. unnaturally large ratio of the
gravity scale ($10^{19}$ GeV) to the electroweak scale ($10^2$
GeV). To solve this problem, theories with large extra dimensions
have been proposed~\cite{Arkani-Hamed:98}. However, they could
only explain a huge value of the Planck mass by introducing
another large scale, namely, a size of extra flat dimensions.
Thus, the hierarchy problem was not really solved, but
reformulated in terms of this new scale.

The model which does solve the problem most economically is the
Randall-Sundrum (RS) model~\cite{Randall:99} with a single extra
dimension and warped background metric~\cite{Kisselev:05}:
\begin{equation}\label{metric}
ds^2 = e^{2 \kappa (\pi r_c - |y|)} \, \eta_{\mu \nu} \, dx^{\mu}
dx^{\nu} + dy^2 \;.
\end{equation}
Here $y = r_c \, \theta$ ($-\pi \leq \theta \leq \pi$), $r_c$
being the ``radius'' of the extra dimension, while $\{x^{\mu}\}$,
$\mu = 0,1,2,3$, are the coordinates in four-dimensional
space-time. The parameter $\kappa$ defines the scalar curvature in
five dimensions. Note that the points $(x^{\mu}, y)$ and
$(x^{\mu}, -y)$ are identified, and the periodicity condition,
$(x^{\mu}, y) = (x_{\mu}, y + 2 \pi r_c)$, is imposed. The tensor
$\eta_{\mu \nu}$ is the Minkowski metric.

It is assumed that there are two 3-dimensional branes with equal
and opposite tensions located at the point $y = 0$ (called the
Plank brane) and point $y =  \pi r_c$ (referred to as the TeV
brane). All SM fields are confined to the TeV brane, while the
gravity propagates in five dimensions. The following relation
between the 4-dimensional (reduced) Planck mass,
$\bar{M}_{\mathrm{Pl}}$, and (reduced) gravity scale in five
dimensions, $\bar{M}_5$, can be derived:
\begin{equation}\label{hierarchy}
\bar{M}_{\mathrm{Pl}}^2 = \frac{\bar{M}_5^3}{\kappa} \left(e^{2
\pi \kappa r_c} - 1 \right) \;.
\end{equation}

The masses of the Kaluza-Klein (KK) graviton excitations are
proportional to the curvature parameter $\kappa$:
\begin{equation}\label{KK_masses}
m_n = x_n \, \kappa , \qquad n=1,2 \ldots \;,
\end{equation}
where $x_n$ are zeros of the Bessel function $J_1(x)$. On the TeV
brane, the zero graviton mode, $h^{(0)}_{\mu \nu}$, and massive
graviton modes, $h^{(n)}_{\mu \nu}$, are coupled to the
energy-momentum tensor of the matter, $T^{\mu \nu}$, as follows:
\begin{equation}\label{Lagrangian}
\mathcal{L}_{\mathrm{int}} = - \frac{1}{\bar{M}_{Pl}} \, T^{\mu
\nu} \, h^{(0)}_{\mu \nu} - \frac{1}{\Lambda_{\pi}} \, T^{\mu \nu}
\, \sum_{n=1}^{\infty} h^{(n)}_{\mu \nu} \;,
\end{equation}
with
\begin{equation}\label{Lambda}
\Lambda_{\pi} = \left( \frac{\bar{M}_5^{3}}{\kappa} \right)^{\!
1/2}
\end{equation}
being a physical scale on this brane.

Let us note that the metric \eqref{metric} differs from that
presented in the original paper~\cite{Randall:99} in which both
$\bar{M}_5$ and $\kappa$ have to be taken as large as the Planck
mass ($\bar{M}_5 \sim \kappa \sim \bar{M}_{\mathrm{Pl}}$).
Moreover, the size of the warped extra dimension should be
extremely small ($r_c \simeq 60 \,
l_{\mathrm{Pl}}$).~\cite{Kisselev:05} Thus, in order to explain
the huge value of $\bar{M}_{\mathrm{Pl}}$ in such a scheme, one
has to introduce new mass scales of the same order, namely,
$\bar{M}_5 $, $\kappa$, and $r_c^{-1}$.

However, the hierarchy problem can be successfully solved in the
RS scenario, but with the metric \eqref{metric}. The equation
\eqref{hierarchy} allows us to consider the {\em small curvature
option} of the RS model~\cite{Kisselev:05}-\cite{Giudice:05}:%
\footnote{For numerical estimates, the region $0.5 \mathrm{\ GeV}
\leqslant \kappa \leqslant 1.5 \mathrm{\ GeV}$ will be used (see
Fig.~\ref{fig:event rate}).}
\begin{equation}\label{small_curv}
\kappa \ll \bar{M}_5 \sim 1 \mathrm{\ TeV} \;.
\end{equation}
In such a case, we get an almost continuous spectrum of low-mass
graviton excitations with the small mass splitting $\Delta m
\simeq \pi \kappa$. Note that in the standard scenario of the RS
model~\cite{Randall:99} one has a series of KK graviton resonances
with the lightest one having a mass around 1 TeV.

The RS model with the large extra dimension has been checked by
the DELPHI Collaboration~\cite{LEP_limit}. The gravity effects
were searched for by studying photon energy spectrum in the
process $e^+e^- \rightarrow \gamma + E_{\perp}\hspace{-6mm}
\diagup \hspace{2mm}$. The limit on $\bar{M}_5$
obtained~\cite{LEP_limit} is
\begin{equation}\label{LEP_limit}
\bar{M}_5 > 0.92 \mathrm{\ TeV} \;.
\end{equation}
The search for large extra dimensions in the diphoton channel
using data collected by the CDF and D\O~Collaborations at
$\sqrt{s} = 1.96$ TeV are presented in Refs.~\cite{Tevatron_ADD}.
The measured $p_{\perp}$-distributions are in a good agreement
with the SM background, that allowed us to obtain the
bound~\cite{Kisselev:08}:
\begin{equation}\label{Tevatron_limit}
\bar{M}_5 > 0.81 \mathrm{\ TeV}\;.
\end{equation}
The discovery limit of the LHC in the two-photon production
(requiring a $5 \sigma$ effect) has also been derived for two
values of the integrated luminosity
$\mathcal{L}$~\cite{Kisselev:08}:
\begin{equation}\label{LHC_limit}
\bar{M}_5 = \left\{
\begin{array}{rl}
  6.3 \mathrm{\ TeV} \;, & \mathcal{L} = 100 \mathrm{\
fb}^{-1} \\
  5.1 \mathrm{\ TeV} \;,  & \mathcal{L} = 30 \mathrm{\
fb}^{-1} \\
\end{array}
\right.
\end{equation}

Previously, the gravity effects in the RS model with the small
curvature were already looked for in a number of different
processes (see Refs.~\cite{Giudice:05}-\cite{Kisselev:08}). In the
present paper we will estimate an upper bound on $\bar{M}_5$ which
can be reached by the Auger ground array in detecting
quasi-horizontal air showers induced by ultra-high energy (UHE)
cosmic neutrinos.

\section{Gravity effects in interactions of cosmic neutrinos
with atmospheric nucleons}

A promising possibility to detect effects induced by the low-mass
KK gravitons \eqref{KK_masses} is to look for their contributions
to the scattering of the SM fields in the trans-Planckian
kinematical region:
\begin{equation}\label{kinem_region}
\sqrt{s} \gtrsim \bar{M}_5 \gg -t \;,
\end{equation}
with $\sqrt{s}$ being the colliding energy and $t = - q_{\bot}^2$
four-dimensional momentum transfer. It is also assumed that
inequality \eqref{small_curv} is satisfied. As we will see below,
in the trans-Planckian region the gravity contribution to the
scattering of UHE cosmic neutrinos off the atmospheric nucleons
can dominate the SM contribution.

In the eikonal approximation which is valid in the kinematical
region \eqref{kinem_region}, elastic scattering amplitude is given
by the sum of gravi-Reggeons, i.e. reggeized gravitons in the
$t$-channel. Because of a presence of extra dimension, the Regge
trajectory of the graviton is splitting into an infinite sequence
of trajectories enumerated by the KK number
$n$~\cite{Kisselev:04}:
\begin{equation}\label{trajectories}
\alpha_n(t) = 2 + \alpha_g' t  - \alpha_g' \, m_n^2, \quad n = 0,
1, \ldots.
\end{equation}
In string theories, the slope of the gravi-Reggeons is universal,
and $\alpha_g' = M_s^{-2}$, where $M_s$ is the string scale. For
more details, see Refs.~\cite{Kisselev:04}.

Correspondingly~\cite{Kisselev:05}, the gravity Born amplitude for
the neutrino scattering off \emph{a point-like}  particle looks
like%
\footnote{Remember that the KK gravitons interact universally with
the SM fields \eqref{Lagrangian}.}
\begin{equation}\label{Born_part_ampl}
A_{\rm grav}^{\mathrm{B}}(s, t) = \frac{\pi \alpha'_g s^2}{2
\Lambda_{\pi}^2} \sum_{n \neq 0} \left[ i - \cot\frac{\pi
\alpha_n(t)}{2} \right] \left( \frac{s}{\bar{M}_5}
\right)^{\alpha_n(t) -2} \;.
\end{equation}

The differential neutrino-proton cross section is of the form:
\begin{equation}\label{diff cross sec}
\frac{d\sigma}{dy} = \frac{1}{16 \pi s} \left| A_{\nu \rm p}(s, t)
\right|^2 \;.
\end{equation}
The inelasticity $y = -t/s$ defines a fraction of the neutrino
energy transferred to the nucleon. $A_{\nu \rm p}$ is the
neutrino-proton amplitude which is related to the eikonal:
\begin{equation}\label{ampl}
A_{\nu \rm p}(s, t) = 4 \pi i \, s \!\! \int\limits_0^{\infty} \!
db b \, J_0(b \, q_{\bot}) \left\{ 1 - \exp [i \chi(s, b)]
\right\} \;.
\end{equation}
In its turn, the eikonal is given by
\begin{equation}\label{eikonal}
\chi(s, b) = \frac{1}{4 \pi s} \! \int\limits_0^{\infty} \!
dq_{\bot} q_{\bot} \, J_0(q_{\bot}  b) \, A_{\nu \rm p}^{\rm B}
(s, t) \;.
\end{equation}
The calculations show that the imaginary part of the eikonal is
negligible with respect to its real part, since $\mathrm{Im}
\chi/\mathrm{Re} \chi = \mathrm{O}(\kappa/\bar{M}_5)$. That is why
we can omit a contribution from inelastic interactions.

The \emph{hadronic} Born amplitude in \eqref{eikonal} is defined
by the gravity amplitude~\eqref{Born_part_ampl}
and skewed ($t$-dependent) parton distributions $F_i(x,t)$:%
\footnote{Since $A_{\rm grav}^{\mathrm{B}} \sim s^2$, the integral
converges rapidly at $x=0$.}
\begin{equation}\label{Born_had_ampl}
A_{\nu \rm p}^{\rm B}(s, t) = \sum_{i=q,\bar{q},g} \int\limits_0^1
dx A_{\rm grav}^{\mathrm{B}}(xs, t) \, F_i(x,t) \;.
\end{equation}
The $t$-dependent distributions have the Regge-like
form~\cite{Kisselev:05}:
\begin{equation}\label{skewed dist}
F_i(x,t) = f_i(x) \exp[t(r_0^2 - \alpha'_{P} \ln x)] \;,
\end{equation}
where $\alpha'_P$ is the Pomeron slope, while $f_i(x)$ is the
distribution of the parton of the type $i$ inside the proton. The
values of the parameters are~\cite{Petrov:02}:
\begin{equation}\label{parameters}
r_0^2 = 0.62 \mathrm{\ GeV}^{-2}, \qquad \alpha'_P = 0.094
\mathrm{\ GeV}^{-2} \;.
\end{equation}
We will use a set of parton distribution functions $f_i(x)$ from
Ref.~\cite{Alekhin:05}.

\begin{figure}[htb]
\begin{center}
\epsfysize=8cm \epsffile{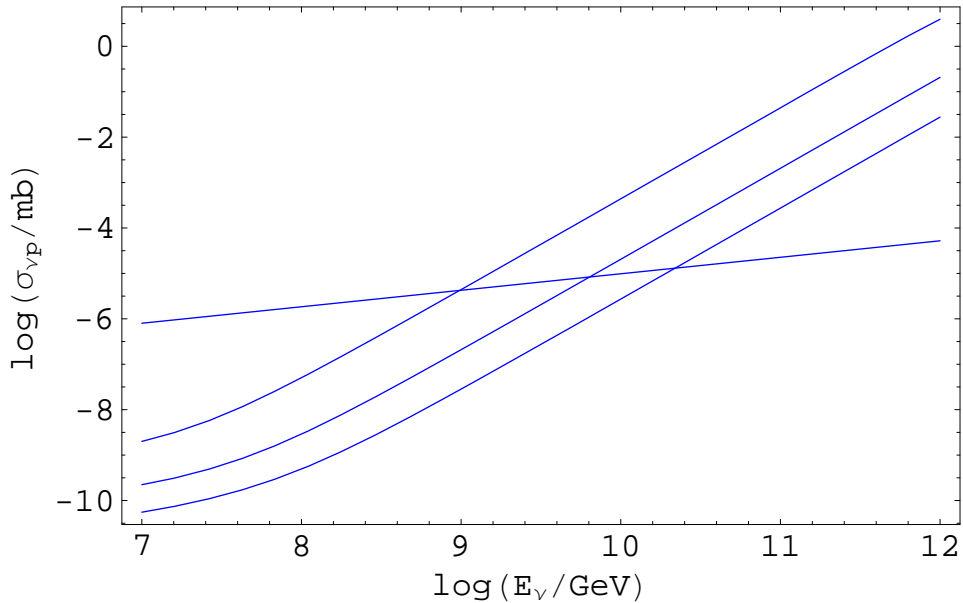}
\end{center}
\caption{The total neutrino-proton cross section as a function of
the neutrino energy (in mb). The curves correspond (from above) to
$\bar{M}_5 = 3$, TeV, 5 TeV, and 7 TeV. The parameter $\kappa$ is
equal to 100 MeV. The straight line: SM neutral current cross
section.}
\label{fig:sigma_100MeV}
\end{figure}

\begin{figure}[htb]
\begin{center}
\epsfysize=8cm \epsffile{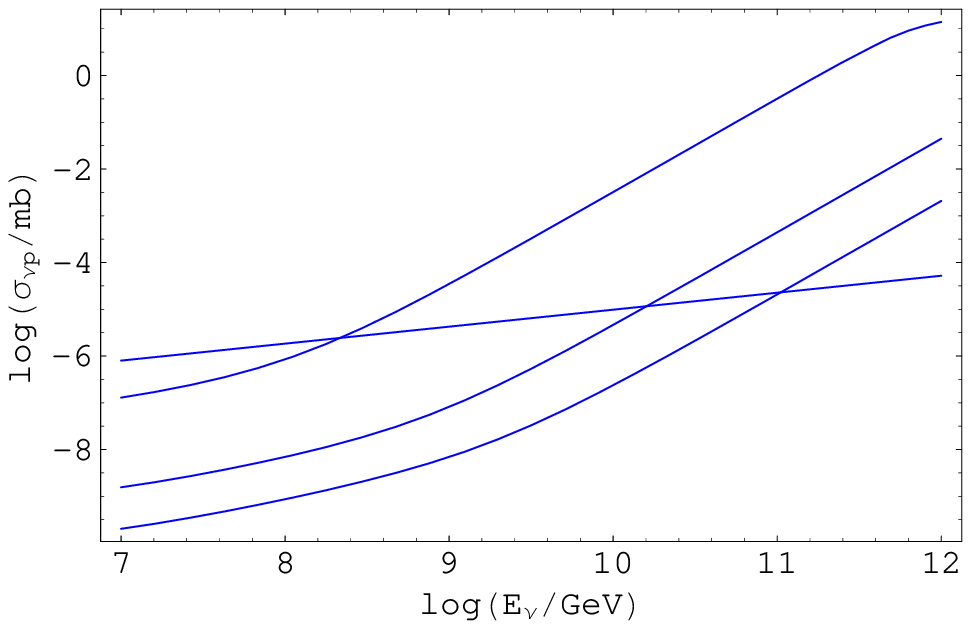}
\end{center}
\caption{The total neutrino-proton cross section as a function of
the neutrino energy (in mb). The curves correspond (from above) to
$\bar{M}_5 = 1$, TeV, 3 TeV, and 5 TeV. The parameter $\kappa$ is
equal to 1 GeV.The straight line: SM neutral current cross
section.}
\label{fig:sigma_1GeV}
\end{figure}

In Fig.~\ref{fig:sigma_100MeV} and Fig.~\ref{fig:sigma_1GeV} we
present total neutrino-nucleon cross sections calculated by using
Eqs.~\eqref{Born_part_ampl}-\eqref{Born_had_ampl} for two values
of the curvature $\kappa$ and different values of the reduced
fundamental gravity scale $\bar{M}_5$.
%%%%%%%%%%%%%%%%%%%%%%%%%%% Fig. 1 %%%%%%%%%%%%%%%%%%%%%%%%%%%%%%%%%%
%%%%%%%%%%%%%%%%%%%%%%%%%%%%%%%%%%%%%%%%%%%%%%%%%%%%%%%%%%%%%%%%%%%%%
%%%%%%%%%%%%%%%%%%%%%%%%%%% Fig. 2 %%%%%%%%%%%%%%%%%%%%%%%%%%%%%%%%%%
%%%%%%%%%%%%%%%%%%%%%%%%%%%%%%%%%%%%%%%%%%%%%%%%%%%%%%%%%%%%%%%%%%%%%

Previously, low-scale gravity effects in cosmic neutrino
interactions were calculated in models with compactified extra
dimensions (see \cite{Anchordoqui:06,Hussain:04} and references
therein). Recently, the gravity effects on the neutrino-nucleon
cross sections in the eikonal approximation were estimated for the
case of infinitely thin branes embedded in five extra
dimensions~\cite{Sessolo:08}. The black hole production cross
sections in cosmic neutrino interactions were also calculated
(see, for instance, Ref.~\cite{Anchordoqui:07}).

By comparing Figs.~\ref{fig:sigma_100MeV}, \ref{fig:sigma_1GeV}
with figures from Ref.~\cite{Sessolo:08}, one can see that the
neutrino cross sections in the small curvature scenario of the RS
model and those in the ADD model have different energy
dependences. The formers are significantly smaller at $E_{\nu}
\lesssim 10^9$ GeV, but exceed the ADD cross sections at $E_{\nu}
\gtrsim 10^{10}$ GeV (at comparable values of gravity scale
$\bar{M}_5$ in both models).

The number of neutrino induced air showers is given by
\begin{equation}\label{event number}
\frac{dN_{\mathrm{ev}}}{dt} = \int_{E_{\mathrm{th}}}^{E_{\max}} \!
dE_{\nu} \! \int_0^1 dy \, \theta (E_{\mathrm{sh}} -
E_{\mathrm{th}}) \, \frac{d \sigma (E_{\nu})}{dy} \, \Phi(E_{\nu})
\, A_{\mathrm{eff}}(E_{\mathrm{sh}}, E_{\nu}) \;,
\end{equation}
where $E_{\nu}$ is the energy of the cosmic neutrino,
$\Phi(E_{\nu})$ denotes its flux, and
\begin{equation}\label{shower_energy}
E_{\mathrm{sh}} = y E_{\nu}
\end{equation}
is the energy of the air shower produced. Effective aperture for
the UHE neutrinos is defined  by the neutrino flux attenuation
$\mathrm{att}(E_{\nu})$ and detector efficiency
$P(E_{\mathrm{sh}})$:
\begin{equation}\label{aperture}
A_{\mathrm{eff}}(E_{\mathrm{sh}}, E_{\nu}) = \mathrm{att}(E_{\nu})
\, P(E_{\mathrm{sh}}) \, A_{\mathrm{p}}(E_{\mathrm{sh}}) \;.
\end{equation}
The attenuation $\mathrm{att}(E_{\nu})$ depends (besides
neutrino-nucleon total cross section) on $X_{\mathrm{obs}}$, the
depth within which air shower is visible for the ground array
detector, and $X_{\mathrm{uno}}$, the minimum atmospheric depth a
neutrino must reach in order to induce an observable shower to
these detectors.
%%%%%%%%%%%%%%%%%%%%%%%%%%% Fig. 3 %%%%%%%%%%%%%%%%%%%%%%%%%%%%%%%%%%
\begin{figure}[htb]
\begin{center}
\epsfysize=10cm \epsffile{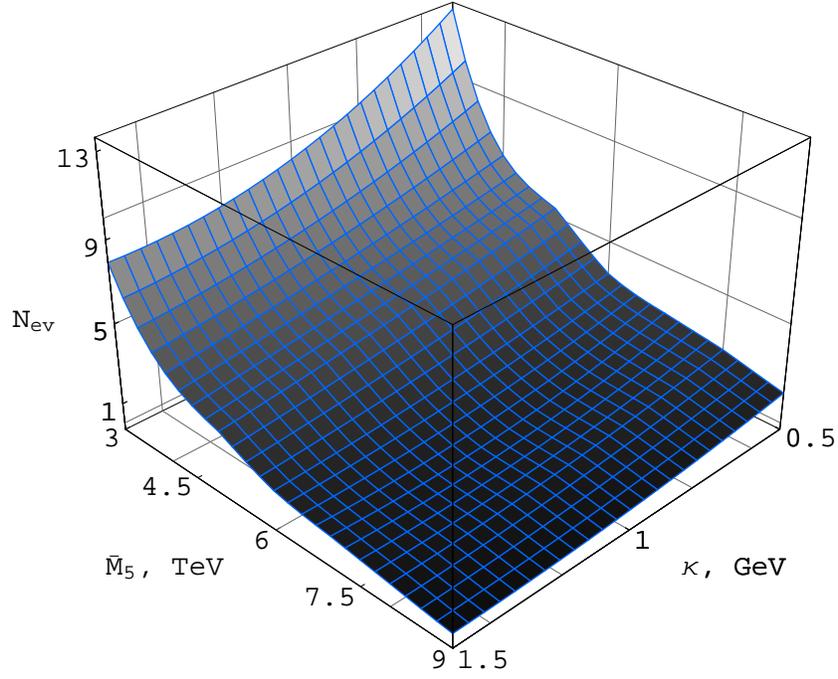}
\end{center}
\caption{The expected rate of the neutrino induced inclined air
showers ($75^\circ \leq \theta_{\mathrm{zenith}} \leq 90^\circ$)
at the Auger Observatory for the Waxman-Bahcall flux (in
yr$^{-1}$).} \label{fig:event rate}
\end{figure}
%%%%%%%%%%%%%%%%%%%%%%%%%%%%%%%%%%%%%%%%%%%%%%%%%%%%%%%%%%%%%%%%%%%%%

In order to isolate neutrino-induced events at the Auger
Observatory, deeply penetrating quasi-horizontal air showers
should be looked for~\cite{Berezinsky:69,Zas:05}. We impose the
following bounds on the zenith angle of the incoming neutrino:
$75^\circ \leq \theta_{\mathrm{zenith}} \leq 90^\circ$. The
functions $P(E_{\mathrm{sh}})$ and
$A_{\mathrm{p}}(E_{\mathrm{sh}})$ as well as values of the
parameters $X_{\mathrm{obs}}$ and $X_{\mathrm{uno}}$ are taken
from Ref.~\cite{Anchordoqui:05}. In particular, the deeply
penetrating events must satisfy the condition $X_{\mathrm{uno}}
\geq 1700$ g/cm$^2$. Since, on average, ultra-high energy air
shower develops
to its maximum after traversing 800 g/cm$^2$,%
\footnote{It corresponds to the criterion $X_{\mathrm{max}} \geq
2500$ g/cm$^2$, where $X_{\mathrm{max}}$  is the shower maximum.}
it is set $X_{\mathrm{obs}} = 1300$ g/cm$^2$ (see
\cite{Zas:05,Anchordoqui:05} for more details).

The threshold energy in \eqref{event number} is taken to be
$E_{\mathrm{th}} = 5 \cdot 10^{7}$ GeV, and maximum energy
$E_{\max} = 10^{12}$ GeV. The result of our calculations for the
Waxman-Bahcall neutrino flux~\cite{WB bound} is presented in
Fig.~\ref{fig:event rate}. It shows the rate of the inclined air
showers at the Auger detector as a function of two parameters of
the model.

In particular, the number of the inclined air showers is equal to
1.54, 0.68, 0.37 for $\bar{M}_5 = 9$ TeV and $\kappa = 0.5$ GeV, 1
GeV, 1.5 GeV, respectively. These estimates can be compared with
the SM prediction, 0.22 events per year for the same neutrino flux
and $\theta_{\mathrm{zenith}} \geq
70^\circ$~\cite{Anchordoqui:06}, that corresponds to $\simeq 0.13$
SM events for our case.

We conclude that the search limit of the Auger Observatory for the
5-dimensional Planck scale $\bar{M}_5$ can reach $9$ TeV
(depending on other parameter $\kappa$), i.e. be even larger than
the discovery limit of the LHC~\eqref{LHC_limit} derived in the
framework of the same scenario.

%%%%%%%%%%%%%%%%%%%%%%%%%%%%%%%%%%%%%%%%%%%%%%%%%%%%%%%%%%%%%%%%%%%%%

%%%%%%%%%%%%%%
% References %
%%%%%%%%%%%%%%

%%%%%%%%%%%%%%%%%%%%%%%%%%%%%%%%%%%%%%%%%%%%%%%%%%%%%%%%%%%%%%%%%%%%%

%%%%%%%%%%%%%%%
\end{document}